# THE CENTRAL REGION IN M100: OBSERVATIONS AND MODELING

J. H. Knapen[1,2], J. E. Beckman[2,3], C. H. Heller[4,5], I. Shlosman[4,6,7]
and R.S. de Jong[8,9]

[1] Département de Physique, Université de Montréal, C.P. 6128, Succursale Centre Ville, Montréal (Québec), H3C 3J7 Canada (present address). E-mail knapen@astro.umontreal.ca
[2] Instituto de Astrofísica de Canarias, E-38200 La Laguna, Tenerife, Spain
[3] E-mail jeb@iac.es
[4] Department of Physics and Astronomy, University of Kentucky, Lexington, KY 40506-0055, USA
[5] Universitäts Sternwarte Göttingen, 11 Geismarlandstrasse, 37083 Göttingen, Germany (present address). E-mail heller@uni-sw.gwdg.de
[6] Gauss Foundation Fellow, Academy of Sciences, Göttingen, Germany
[7] E-mail shlosman@asta.pa.uky.edu
[8] University of Durham, Department of Physics, South Road, Durham DH1 3LE, United Kingdom (present address). E-mail R.S.deJong@Durham.ac.uk
[9] Kapteyn Astronomical Institute, Postbus 800,
NL-9700 AV Groningen, the Netherlands. E-mail roelof@astro.rug.nl

## Abstract

We present new high-resolution observations of the central region of the late-type spiral galaxy M100 (NGC 4321) supplemented by 3D numerical modeling of stellar and gas dynamics, including star formation (SF). Near-infrared imaging has revealed a small bulge of $4''$ effective diameter; a $60''$ radial length stellar bar of moderate strength, previously inferred from optical and 21 cm observations; and an ovally-shaped, ring-like structure in the plane of the disk between $10''-22''$ from the center, whose major axis makes a large angle with the bar. The $K$ isophotes become progressively elongated and skewed to the position angle of the bar *both* outside and inside the "ring", forming an inner bar-like region. The galaxy exhibits a mild circumnuclear starburst concentrated in the inner part of the $K$ "ring". This SF is prominent in H$\alpha$ and the $U, B$ and $V$ bands, forming an incomplete ring. In addition, two strong local maxima of $K$ emission have been observed to lie remarkably symmetrically with respect to the galactic nucleus and equidistant from it at $7\rlap{.}''5$ slightly leading the stellar bar. CO molecular emission is peaked in the dust lanes seen in the $I-K$ color index image.

We interpret the twists in $K$ isophotes and the swinging of spiral arms through $\sim 360°$ inside the corotation radius as being indicative of the presence of a double inner Lindblad resonance (ILR) and test this hypothesis by modeling the gas flow in a self-consistent gas+stars disk embedded in a halo, with an overall NGC 4321-like mass distribution in the system. Both ILRs have been verified using nonlinear orbit analysis and by determining the spatial extent of the family of orbits oriented along the minor axis of the bar.

We have reproduced the basic morphology of the region inside corotation, including (1) the $\sim 60''$ bar; (2) the large scale trailing shocks representing the offset dust lanes in the bar; (3) two symmetric $K$ peaks corresponding to gas compression maxima which lie at the caustic formed by the interaction of a pair of trailing and leading shocks in the vicinity of the inner ILR, both peaks being sites of SF; and (4) two additional zones of SF corresponding to gas compression maxima at the bar's minor axis, where the large scale shocks start to curl and which are referred in the literature as "twin peaks". We further argue that the twisting of $K$ isophotes in the neighborhood of the resonance region requires a population of red stars which are dynamically young and follow gas rather than stellar orbits, *i.e.* red supergiants. At the same time a substantial contribution from OB stars to the $K$ light is expected within the inner kpc and especially in the symmetric $K$ peaks. We also conclude that NGC 4321 hosts a single stellar bar which fuels the starburst activity within the circumnuclear "ring" by channeling gas there at the median rate of $\sim 0.1 - 1 \, M_\odot \, yr^{-1}$.

*Subject headings:* galaxies: individual (M100, NGC 4321) — galaxies: kinematics & dynamics — galaxies: nuclei — galaxies: evolution, internal motions, structure — galaxies: starburst

# 1  Introduction

The central regions of disk galaxies are frequent sites of enhanced activity, most probably fueled by the inflow of galactic interstellar medium (hereafter ISM). Known manifestations, Seyfert nuclei and nuclear starbursts, differ substantially in their physical nature and activity level, which in most cases cannot be maintained for periods much longer than $\sim 10^8$ yrs without some refueling mechanism (*e.g.* reviews by Begelman, Blandford & Rees 1984; Telesco 1988; Heckman 1990; Shlosman, Begelman & Frank 1990; Athanassoula 1994; and refs. therein). This must entail a means of resupplying gas used up in stars and lost in galactic superwinds driven by nuclear starbursts, or accreted by central black holes (BHs) in active galactic nuclei (AGNs).

The problem of driving the gas into the center of a disk galaxy is that of shedding angular momentum in a rotating system. On galactic scales, $\sim 1 - 10$ kpc, this can be achieved by means of gravitational torques from non-axisymmetric features such as stellar bars, oval distortions, and spiral arms, as suggested originally by Simkin, Su & Schwarz (1980). Gravitational torques result in a catastrophic loss of angular momentum by the ISM; the efficiency of this process depends on the particular perturbation and its amplitude. The gas dynamical processes become increasingly important on scales $\lesssim 1$ kpc, as noted by Shlosman, Frank & Begelman (1988, 1989) and Heller & Shlosman (1994), where (1) the gas gravity can affect the stellar dynamics, inflow rates and star formation (SF); (2) inner Lindblad resonances (ILRs) between a stellar disk and a bar, if present, are expected to be found; and (3) a radial inflow should slow down in the neighborhood of ILRs (or the $Q$-barrier, where $Q$ is Toomre's [1964] parameter), and the resulting accumulation of gas may serve as a precursor of a nuclear starburst. Finally, deep within the potential well and a few pc from the BHs in AGNs, an angular momentum transport may be dominated by magnetic torques (Blandford & Payne 1982; Emmering, Blandford & Shlosman 1992).

Departure from axisymmetry on the galactic scale may be very explicit in some cases: *e.g.* about half of the observed disk galaxies are barred (*e.g.* de Vaucouleurs 1963; Sellwood & Wilkinson 1993 and refs. therein); or less evident in other: many galactic disks are ovally distorted which is clear both on kinematic (Bosma 1981) and photometric (Kormendy 1982) grounds; and some bulges are triaxial (Norman & Silk 1983; Pfenniger & Norman 1990; Kormendy 1994). In fact, as more near-infrared (NIR) images of galaxies become available, an increasing number of galaxies are seen to have major distortions and bars in their stellar mass distribution (*e.g.* Hackwell & Schweizer 1983; Scoville *et al.* 1988; Thronson *et al.* 1989; Telesco *et al.* 1991; Zaritsky, Rix & Rieke 1993; Block *et al.* 1994). Thus the basic dynamical precondition for sustained circumnuclear activity at some level appears to exist in the majority of disks. On the other hand, observational evidence supporting the recent redistribution of the ISM on galactic scale and correlation of central activity with the presence of non-axisymmetric features is growing (Lo *et al.* 1987; Nakai *et al.* 1987; Ishizuki *et al.* 1990; Meixner *et al.* 1990; Taniguchi *et al.* 1990; Wang, Scoville & Sanders 1991; Kenney *et al.* 1992; Martin & Roy 1994; Scoville *et al.* 1994; Moles, Márquez & Pérez 1995).

Central starburst activity is enhanced in barred spirals, as evidenced from optical, H$\alpha$, radio and infrared (IR) surveys (Adams 1977; Simkin, Su & Schwarz 1980; Hummel 1981; Balzano 1983; Devereux 1987; Puxley, Hawarden & Mountain 1988; Telesco, Dressel & Wolstencroft 1993; see also review by Kennicutt 1994 and refs. therein). It often delineates (roughly) kpc-size ring-like structures of SF regions mixed with dust, called nuclear rings. At superior resolution these "rings" may consist of tightly wound spiral arms, be patchy and incomplete (*e.g.* Pogge 1989; Buta & Crocker 1993). Rings have been detected in UV, optical, NIR and radio continua, as well as in H$\alpha$ and CO, and are associated with the ongoing SF. In early-type galaxies, S0/a — Sab, the SF activity, as manifested by bright H II regions, is almost entirely limited to these rings (*e.g.* Pogge 1989; Dressel & Gallagher 1994). In late-type spirals, the nuclear rings also stand out, due to their high surface brightness.

The position and morphology of nuclear "rings" may be directly related to the large scale dynamics in disk galaxies for which they in fact serve as probes. In particular, they seem to be associated with the ILRs of barred stellar disks and as such impose constraints on the bar pattern speed and on the possible gas flow (*e.g.* Telesco & Decher 1988; Kenney *et al.* 1992; Athanassoula 1994). Moreover, nuclear "rings" correlate with the presence of leading dust lanes, which themselves delineate offset shocks in the stellar bar (Prendergast 1962; Athanassoula 1992). Properties of nuclear "rings" have been studied in a number of works, observationally (Buta 1986a,b; Buta & Crocker 1993; Shaw *et al.* 1993; and others) and theoretically (Schwarz 1984; Combes & Gerin 1985; Athanassoula 1992; Elmegreen 1994; Heller & Shlosman 1995). Their intrinsic shapes vary from circular to moderately elliptical and they lead the stellar bars typically by $\sim 50° - 90°$.

The galaxy M100 (NGC 4321), the brightest spiral in the Virgo cluster, has been known to exhibit mild SF activity in the circumnuclear region since work of Morgan (1958) and its in-clusion in the Sérsic & Pastoriza (1967) list of "hot spot" galaxies. The spectrum, dominated by H II regions in the "ring", changes to LINER-type when using smaller apertures, thus showing that the nucleus is probably mildly active (Kennicutt, Keel & Blaha 1989). Classified by Sandage (1961) as Sbc, NGC 4321 is favorably inclined at $30° ± 3°$ (de Vaucouleurs *et al.* 1976 [RC2], see also Knapen *et al.* 1993a), and its barred nature has been noted in the RC2 where it was defined as an SAB(s)bc. A pair of offset dust lanes penetrating the central region can be clearly observed in an optical photograph (Sandage & Bedke 1988). Both the stellar bar and the central oval distortion, associated with the 1 kpc ring of SF regions, have been confirmed and studied by Pierce (1986) and Arsenault *et al.* (1988). Pierce argued in favor of an ILR in the vicinity of the ring, Arsenault *et al.* based their claim of two ILRs on the H$\alpha$ rotation curve. The "ring" has a distinct four-arm spiral pattern (Pogge 1989; Cepa & Beckman 1990). Knapen *et al.* (1993a) confirmed the few kpc bar using a broad *H*-band image in the NIR (shown in Fig. 5 of the present paper), and by constructing a velocity field of the 21 cm H I emission. This revealed gas flow in highly elongated orbits along the bar major axis, in accordance with hydrodynamical models (*e.g.* Huntley 1980; Athanassoula 1992; Friedli & Benz 1993; Heller & Shlosman 1994). Finally, Knapen *et al.* (1995a; Paper I) have studied optical and NIR morphologies of the central kpc region at high resolution exposing the inner bar-like region. The galaxy has also a small companion NGC 4322, which may be causing some distortion and asymmetry in the gas velocity and distribution of the outer disk, as picked out especially at 21 cm (Knapen *et al.* 1993a). In the present paper we use the distance of $D = 13.6$ Mpc to NGC 4321 (de Vaucouleurs 1984), thus $15''$ corresponds to 1 kpc and $1''$ to 67 pc. (This distance was recently updated to 17.1 Mpc [Freedman *et al.* 1994], but as our numerical simulations are not scalable, we retain the old value without the loss of generality.)

This paper is devoted to observational and numerical studies of gas and stellar responses within the resonant central region of a barred disk galaxy with a particular emphasis on NGC 4321. We aim at understanding the observed morphology and its relationship to the SF processes there. The numerical part, especially that related to modeling of SF, should be viewed as providing qualitative insight into the evolution. In Section 2 we present high resolution H$\alpha$ and NIR *K*-band photometric images of the central region in NGC 4321 supplemented with broad-band optical observations. In Section 3 we examine in detail the morphology of the stellar populations and that of the gas as inferred from the dust and CO distributions (the CO data were kindly provided by B. Canzian). This is followed (Section 4) by modeling numerically the evolution of the inner part of a two-component, gas+stars, self-gravitating disk whose rotation curve follows from an NGC 4321-like radial mass distribution, focusing on the formation of nuclear rings. In Section 5, we show that the observed morphology in the central kpc can be explained by resonant response to the perturbing force of a moderate strength stellar bar in the presence of two ILRs. Additional comments are given in the last section.

## 2 Observations

### 2.1 Optical and NIR Observations

Broad-band optical images in the $U, B, V, R$ and $I$ Johnson photometric bands were obtained the night of May 14, 1993, at the 4.2m William Herschel Telescope (WHT) at La Palma, using the Auxiliary focus camera at the $f/11$ Cassegrain focus of the telescope. An EEV P88300 CCD detector was used, which gives an image scale of $0\rlap{.}''10$ per $22\mu$m CCD pixel, with a field of view of $2'$ diameter. The images were reduced in a standard way removing a bias level, correcting large scale variations in sensitivity over the chip surface using dawn flatfield images, and subtracting the sky background level using sky frames taken of a nearby empty field shortly after the galaxy images. Since the exposure times were short, the number of cosmic rays on the images is small and no attempts were made to remove them. Exposure times were 600 s for the image in the $U$ band, 180 s for that in $B$, and 60 s for those in $V, R$ and $I$. Seeing values were $0\rlap{.}''8 \pm 0\rlap{.}''1$ FWHM. The images were taken under photometric conditions, and calibrated by observing a standard star from the list of Landolt (1983).

A $K$-band image ($\lambda = 2.2\mu$m; $\Delta\lambda = 0.42$ $\mu$m) was obtained by R.F. Peletier on the night of June 5th, 1994 at the UK IR Telescope (UKIRT) using the IRCAM3 camera (Paper I). This camera was equipped with a $256 \times 256$ array, giving a pixelscale of $0\rlap{.}''286$ and a field of view of about $70'' \times 70''$. Twelve separate images of the central region in NGC 4321, 60 s each, interleaved with 8 exposures of the sky background were taken. The standard data reduction procedures of bias subtraction, flatfielding (using the night sky frames) and averaging were applied to produce the final frame, a sum of the four separate images with best seeing. The resolution of the final image is some $0\rlap{.}''8$. The image was calibrated by comparing it with a photometric $K$-band image obtained earlier (de Jong & van der Kruit 1994).

The images in the different photometric bands were aligned by fitting Gaussian profiles to a number of field stars in the images. We also obtained the exact resolution from these fits. This positioning was performed to a precision of better than 0.2 of a pixel, or $0\rlap{.}''02$ ($0\rlap{.}''05$ in $K$), much better than the resolution in the images. The positions of some of the stars, as determined using the HST Guide Star Catalog, and the position of the center of the galaxy were used to place the images on a correct R.A.—declination grid, with an estimated precision of a few arcsec. The central $40'' \times 40''$ area of the images in the $U, B, V, R, I$ and $K$ bands are shown in Fig. 1. A more detailed $K$-image in contour representation is given in Fig. 2.

Color index images have been produced by dividing images in different photometric bands. To take account of the slightly different resolutions of the individual images, for each pair of images we convolved the one with the better seeing with a Gaussian to yield the resolution of the other image. The $U - I$ image, with a resolution of $0\rlap{.}''85 \times 0\rlap{.}''85$ is presented in Fig. 3. An optical-NIR $I - K$ color index image (made after transforming the $K$-band image to an artificial grid of $0\rlap{.}''1$ pixels), also with a resolution of some $0\rlap{.}''8$ FWHM, is shown in Fig. 4.

## 2.2 Narrow-band H$\alpha$ Observations

The H$\alpha$ line observations of NGC 4321 were made in May 1991 (eastern half of the galaxy) and March 1992 (western half) using the 4.2 m WHT with the TAURUS instrument in imaging mode, during bright and photometric sky conditions. An EEV CCD chip with pixels of projected size $0''\!.27 \times 0''\!.27$, giving a field of view (vignetted by the filter) of some $5'$ diameter was used. Redshifted 15Å wide narrow band interference filters were applied, both for the line (using a 6601Å filter) and the continuum images (6577Å and 6565Å filters for eastern and western half, respectively). Exposure times were 1200 s for both the on-line and the continuum images of the eastern half, and $2 \times 900$ s for the on-line and continuum images of the western half of the galaxy.

Since the galaxy is larger than the available field of view, it was observed in four different fields, making sure the center of the galaxy was well visible in a corner of the chip. For each of these images, the basic reduction procedure (flatfielding, positioning of the separate images, continuum subtraction and calibration) was followed as described by Knapen *et al.* (1993b) for the H$\alpha$ image of NGC 6814. The four continuum-subtracted H$\alpha$ frames were then scaled and combined into a mosaic using positions of the galaxy center and some foreground stars seen in at least two different frames. The exact RA-dec orientation was determined by measuring the positions of foreground stars in the image that are listed in the HST Guide Star Catalog. The resulting total image covers all of the disk, up to well outside $D_{25}$, with a resolution (seeing FWHM) of less than $1''\!.0$ over the whole field. Its inner part is shown in Fig. 5. The inset in Fig. 5 was taken from the sub-image that had the best resolution, seeing is $\sim 0''\!.8$.

## 2.3 Hubble Space Telescope observations

The HST was used to image the central region of NGC 4321 employing the restored WFPC. One of these images was retrieved from the HST archive: a 900 s exposure obtained on Dec. 31st, 1993 with the planetary camera through the F555W filter (roughly equivalent in wavelength to a $V$ band filter). A considerable number of cosmic ray hits were removed from the image by replacing affected pixels by values interpolated from neighboring pixels. The resolution of the image as used here is considerably higher than that of all the other data. The HST image is shown in comparison with the $K$ image (which is overlaid in contours) in Fig. 6.

## 2.4 CO interferometric data

For comparison with the optical and NIR data, we used a CO emission image obtained by Canzian (1990) in 1986 with the Owens Valley mm-wave interferometer array. The observations, calibration, and further reduction are described in detail by Canzian (1990). The $65''$ primary beam covers the inner region of the galaxy, and the synthesized beam in the image as used here is $6''\!.9 \times 4''\!.4$ with a position angle (PA) of $-3.5°$. Positional accuracy is estimated to be one fifth of the beam, or $\sim 1''$. The CO contour map is shown in Fig. 7, overlaid on the $I - K$ color index map of Fig. 4.

## 3 Observed morphology

### 3.1 Disk morphology

Understanding of the circumnuclear morphology in NGC 4321 relies on its relationship to the rest of the galaxy which is a grand-design two-armed spiral (see Peletier 1994 for a high quality true-color image). Elmegreen & Elmegreen (1984) classify the galaxy in their "class 12", *i.e.* having very well-defined arms, although the symmetry of its recent SF is less striking than that of M51 (Knapen *et al.* 1992; Knapen *et al.* 1995b). The massive SF efficiency in the arms is some three times higher than in the interarm disk, in the rather well-defined southern arm, and also in the more flocculent northern arm (Knapen *et al.* 1995b). The IR-to-blue luminosity for NGC 4321 is $L_{FIR(1-500\mu m)}/L_B = 0.42$ (Young *et al.* 1989), indicating this galaxy is relatively quiet, compared to *e.g.* AGN hosts or extreme starbursts. The galaxy contains a bar, of radial length 4 kpc at PA= $107° \pm 3°$, close to the PA found by Pierce (1986), using an $I$-band CCD image, and confirmed by Knapen *et al.* (1993a) using a combination of H I (21 cm) kinematics and NIR ($H$-band) photometry. This is reproduced in Fig. 5, where the H isophotes are strongly elongated at the PA of the bar, and where clear H$\alpha$ peaks are visible near its ends. Little SF is observed along the bar at smaller radii until the circumnuclear zone is reached at $\sim 500 - 1,400$ pc from the center. Inside $\sim 500$ pc the SF drops again (with the possible exception of the nucleus). As seen in Fig. 5, the only SF associated with the bar, apart from at its ends, shows up to the NW of the center as a string of H II regions. These regions lie just north of the major dust lane seen in optical images. The spiral arm which they outline connects the main disk of the galaxy, through the end of the bar, to the central kpc region indicating a dynamical connection to the rest of the galaxy. The same pattern can be seen on the SE side of the bar by following the dust lane, but it shows up less clearly here in H$\alpha$.

Dust lanes are visible throughout the disk, but a detailed study of scale-lengths through broad-band filters $(B, V, R, I)$ shows that on average the optical depth in dust is low, both in the arms and in the interarm zones (Beckman *et al.* 1995). The corotation radius in the disk has been estimated to be at $\sim 5$ kpc, using an ionized gas rotation curve (Arsenault *et al.* 1988), and at $5 \pm 0.5$ kpc (Canzian & Allen 1995), following the 2D test for reversal of streaming motions (Canzian 1993) in the H$\alpha$ velocity field. Sempere *et al.* (1995), however, find a corotation radius of some $7.3 \pm 0.7$ kpc, well beyond the ends of the bar, applying a similar test to an H I velocity field, a value in agreement with the results of the morphological determination by Elmegreen, Elmegreen & Seiden (1989).

In the outer parts of the H I disk Knapen *et al.* (1993a) found a deviation in the velocity field accompanied by a faint extension in H I density, most probably due to the companion NGC 4322. Its small size and a relatively large separation from the parent galaxy speak against a strongly interacting system NGC 4321/4322. We hence neglect the effect of NGC 4322 on the SF and dynamical evolution in the central kpc of NGC 4321.

## 3.2 Optical and NIR morphology of the central zone

Paper I described the most important morphological features in our $K$ band and optical images. Here we summarize this discussion and provide additional details which were left out.

The H$\alpha$ morphology around the nucleus (Fig. 5) is dominated by four main components (labeled H$\alpha$1 to H$\alpha$4), first picked out by Pogge (1989), and tentatively identified by Pogge and by Cepa & Beckman (1990) as a four-armed mini-spiral. Each component is broken up into a number of individual "hot spots" where strong SF is taking place, and which are also clearly seen in the $U$- and $B$-band images (Fig. 1). 16% of all the H$\alpha$ emitted by NGC 4321 comes from this zone of some $30''$ diameter, yielding $\sim 3.0 \pm 0.5 \times 10^{40}$ erg s$^{-1}$. The nucleus, although clearly detected in H$\alpha$, is not particularly strong, and shows no measured extension. Inside the area bracketed by the four H$\alpha$ components, a region of reduced H$\alpha$ intensity is seen (inset Fig. 5), whose isophotes of minimum H$\alpha$ emission lie along the general direction of the bar.

Two color index images of the circumnuclear region give characteristic information on stars and dust, respectively. In the $U-I$ image (Fig. 3), darker parts correspond to redder regions in the galaxy, which can be caused by dust extinction, or by a change in stellar population (young stars causing more blue emission). It is not possible to distinguish between these two effects on the basis of a $U-I$ map alone. In the $I-K$ color index map (Fig. 4) there is relatively little change due to varying stellar populations between $I$ and $K$ (Rix & Rieke 1993), but while the effects of dust extinction are low in $K$ ($A_K/A_V = 0.11$), they are quite substantial in $I$ ($A_I/A_V = 0.48$). A red (or dark in Fig. 4) zone in the $I-K$ map thus directly indicates the presence of dust there. It reveals a picture of two dust lanes coming in from outside the inner 1 kpc zone along the concave edges of the spiral arms in the disk, swinging in towards the inner bar-like region, and continuing further on, practically parallel to the bar major axis. The $U-I$ map defines the detailed locations of SF. In Fig. 3, the pattern of four elongated segments mentioned above can be seen clearly, together with the smaller "hot spots" which compose each segment. From the $I-K$ (Fig. 4) and HST $V$-band (Fig. 6) images it is clear that the inner spiral is in fact two-armed, not four-armed as indicated by the blue and H$\alpha$ emission, and most clearly delineated by the strong dust lanes running along the arms. The four-segmented appearance is due to dust lanes crossing the arms, most clearly seen to the SE at $\sim 8''$ from the nucleus, and enhanced SF at specific places along the arms.

In order to study the detailed morphology in the NIR, we determined the run against radius of the azimuthally averaged values of surface brightness ($\mu$, in mag arcsec$^{-2}$), PA and ellipticity ($\epsilon$, defined as $1 - b/a$, where $b/a$ is minor-to-major axis ratio) of the $K$-image ($R < 35''$) and of an $I$-image of the whole disk of the galaxy (used for $10'' < R < 80''$). Ellipses were fitted to the isophotes, making a least squares fit to the pixels in a selected intensity range in the image. Unwanted pixel values due to foreground stars and cosmic rays were removed interactively. In order to check the consistency of the fits, each ellipse was viewed overlayed on the original image, with the selected pixels highlighted. In most cases ellipses were fitted at 0.2 mag arcsec$^{-2}$ spacings. Although the results from the two images can be compared in terms of structural properties of the disk, it must be kept in mind that properties of dust and stellar populations differ from $I$ to $K$, and that results from the two passbands are not interchangeable. The results are presented in Fig. 8, showing the run of the $K$ and $I$ surface brightness, PA, and $\epsilon$ against radius.

The approximate range in radius of the separate components, as named in Paper I, where most of them were morphologically defined from the $K$-image, is indicated in the middle panel of Fig. 8. These components are, from inside out: i) a small *bulge* component ($0'' < R < 2''$; $\epsilon < 0.2$); ii) the *inner barlike region* ($2'' < R < 4''.5$; PA=$111° \pm 1°$; $\epsilon$ slowly rising from 0.2 to 0.5); iii) two *leading arms* ($4''.5 < R < 7'' - 8''$; $\epsilon$ is reaching a maximal value of $0.6 \pm 0.04$) in a region where the contours in the $K$-image clearly deviate from what is expected from a bar by becoming progressively twisted in the direction of rotation, towards larger PA (anti-clockwise), *i.e.* in the leading sense with respect to the bar. This twisting occurs in the region where leading arms are expected theoretically and are modeled numerically by us (Sect. 5); iv) two *peaks of emission* (K1 and K2 in Fig. 3) ($R \sim 7'' - 9''$; PA of the line joining these peaks is $114° \pm 1°$); v) the oval *ring-like* zone where the SF armlets are found in the optical and H$\alpha$ ($10'' < R < 22''$; maximum PA=$159° \pm 3°$ at $R \sim 20''$; minimum $\epsilon = 0.13 \pm 0.02$ near $R = 17 - 18''$). The $K$ contours in this region continuously twist in the leading direction, are smooth, and show hardly any indication of spiral arms or dust lanes. As we argue below (Sect. 6), these apparent changes in morphology are caused by population changes, along with the much reduced absorption by dust at $2.2\mu$; vi) the *large scale stellar bar* ($R > 30''$; PA$\approx 107° \pm 3°$ outside $R = 35''$; maximum $\epsilon = 0.52 \pm 0.04$); vii) the *main disk* component (not shown in Fig. 8; $R > 90''$; PA= $153° \pm 2°$ and ellipticity $\epsilon = 0.13 \pm 0.03$ take the main disk values [Knapen *et al.* 1993a]).

The surface brightness profile shows a relatively steep slope in the region of the ring-like SF zone, and a more shallow decrease in the regions of the two bar components (inner and large-scale) surrounding it. In the main disk, the surface brightness shows an exponentially falling profile (see *e.g.* Beckman *et al.* 1995).

The determinations of PA and ellipticity indicate that the zone where the SF armlets are seen in H$\alpha$ ($10'' < R < 22''$) most probably is intrinsically round, and lying in the plane of the main disk. The inner barlike structure and the large-scale bar have the same PA and similar ellipticity and are thus aligned to within the errors of the fitting procedure. Both are of roughly the same strength — a further indication that only one stellar bar is present. The gradual skewing of $K$ and $I$ isophotes inside and outside the SF zone so notable in Fig. 8, in PA and $\epsilon$, is an indication for the presence of ILRs, and will prove to be important when we discuss the interplay of dynamical resonances with the gas flow and SF.

The $K$ peaks K1 and K2 are elongated perpendicularly to the bar's major axis, for 2-3'', along the spiral arm segments seen in H$\alpha$ (labeled H$\alpha$1 and H$\alpha$2 in Fig. 5). The strongest H$\alpha$ peak in the H$\alpha$1 complex corresponds in position to $K$-peak K1, whereas the strongest H$\alpha$ peak in the H$\alpha$2 complex is offset from the K2 peak by some $1''.5$. The H$\alpha$ peak coinciding with K1 is at the position where the HST image (Fig. 6) and the $I - K$ color index image (Fig. 4) show strong emission surrounded by a dust shell. There is no obvious H$\alpha$ counterpart centered on the K2 peak although the H$\alpha$2 complex surrounds K2 and is clearly associated with it. Paper I showed that the $UVK$ colors of selected regions, combined with stellar population models, are compatible with a scenario in which the four peaks K1, K2, H$\alpha$3 and H$\alpha$4 are caused by coeval bursts of SF, but that whereas H$\alpha$3 has somewhat less dust than the similar regions H$\alpha$4 and K1, light from K2 is much more reddened by dust extinction and/or absorption. We conjecture that K1 and K2 are two similar starbursts, where K1 has already broken through the dust, whereas K2 is still buried in the dust lane.

## 3.3 Effects of dust extinction at 2.2$\mu$m

The $I - K$ color image was used to make a first-order correction for presence of localized dust extinction in the $K$-image (Paper I). We concluded that the observed morphology at 2.2$\mu$m is barely influenced by dust. In correcting the $K$-image, two assumptions have been made, which are discussed explicitly here.

First, we assumed that the $I - K$ color of the stars in the central region ($R < 15''$) is constant. This assumption is not very severe, and invokes the property that $I - K$ remains remarkably constant for stellar populations older than $\sim 10^8$ yrs (see the models of Bressan, Chiosi & Fagotto 1994). Apart from the most recent bursts of SF, which have lower (bluer) $I - K$, and show up as white patches in Fig. 4, we can assume that the intrinsic $I - K$ color in most of the central region of NGC 4321 lies between 1.65 and 2.0. Since $\sim 90\%$ of the pixels in our $I - K$ image take values between 1.7 and 2.5, the extinction E($I - K$) here is never more than 0.85 mag, implying that we are in the optically thin regime in $I$ over virtually the whole central area.

Second, we assumed a Galactic extinction law (Rieke & Lebovsky 1985) in deriving A$_K$ from E($I - K$). Knapen *et al.* (1991) and Jansen *et al.* (1994) verified the applicability of this law in a number of external galaxies, showing in particular that geometrical effects, including the use of a pure screen model with all the dust in front of the stars, leads to only small differences, especially in the red and NIR parts of the spectrum. Since we are interested in an order-of-magnitude estimate of the extinction in $K$, this assumption is justified. Using the Galactic extinction law the extreme value of E($I - K$) = 0.85 would correspond to A$_K$ = 0.28 mag, or about one contour in Fig. 2, implying that indeed our extinction correction is very small.

The above assumptions lead to a small correction to the $K$ morphology due to dust extinction. Paper I showed that such important features as the leading arms and zones K1 and K2 are not artifacts of dust obscuration (in fact, the leading arms seem to be somewhat hidden by dust). An additional argument in favor of only a small dust correction in $K$ is that a patch of extinction in $K$ of, say, 0.5 mag has a corresponding extinction in $V$ of some $2 - 4$ mag, and even larger in $U$ ($\sim 3 - 6$ mag). Such values are excluded by the morphology of the individual and color index images (Figs. 1, 3), as well as by the surface brightness profile of NGC 4321, which is very much within the normal for spirals of its type (*e.g.* de Jong & van der Kruit 1994). We thus feel confident that the morphology as seen in the $K$-band image is not significantly influenced by localized dust absorption, and that we can use this image as a solid basis for our modeling and interpretation.

### 3.4 CO

In Fig. 7 we show a contour representation of the CO interferometric map of the central region overlaid on a grey-scale map of the $I - K$ color index image. The CO traces localized concentrations of molecular hydrogen gas, but due to the lack of short spacings in the observations is not sensitive to structure on a scale larger than $30''$, or maybe even $20''$ (B. Canzian, private communication). From single dish observations by Kenney & Young (1988), Cepa *et al.* (1992) and García-Burillo *et al.* (1994) it is known that the CO distribution in this galaxy rises roughly exponentially towards the center, with an additional central component, possibly due to the bar (Knapen *et al.* 1995b). We do *not* see the bulk of this more slowly varying emission in the interferometric image presented here, but can use the latter map to trace concentrations in the molecular gas that may be due to shocks. At a distance of some $5''$–$10''$ four main concentrations are seen in CO, of which the ones to the North and South of the center are double. Only the largest of these, to the North, seems resolved at the $6''\!.9 \times 4''\!.4$ resolution of the map, the five others (including the one at the center) are most probably unresolved. The appearance of the CO map is generally confirmed by a new BIMA map with higher sensitivity and resolution discussed by Rand (1995), although the smallest of the two peaks to the North is not present, and the peak West is less pronounced in Rand's data. These differences may be due to differences between the telescope arrays used (Rand 1995).

Comparing the positions of the circumnuclear CO concentrations with the $I - K$ color index map, where dust shows up as darker regions, it is clear that all CO peaks are positioned in regions of high dust extinction, the North and South concentrations lying in the main dust lanes, where strongest shocks are found in our numerical models. The peaks toward the East and West occur at the end of the inner barlike feature, where gas orbits crowd and shocks are expected (see below). The CO peaks are also closely related to the main H$\alpha$ peaks, although lying slightly outside in all cases. We infer that the four CO peaks must correspond to strong molecular concentrations, but only better resolved data, preferably probing different transitions or even molecules, plus modeling could reveal to what extent these peak brightness temperatures are due to heat input from the SF regions, and what is the precise physical relation between these quantities.

## 4 Modeling

To understand more fully the response of the gas to the underlying gravitational potential of the central region in NGC 4321 and especially the formation of the nuclear ring-like structure and associated features, we have modeled the stellar and gas dynamical processes there by means of 3D numerical simulations using a method described by Heller & Shlosman (1994). Here we repeat only the necessary details and discuss modifications.

The time-dependent gravitational interactions between gas and stars have been calculated using the TREE algorithm (Barnes & Hut 1986; Hernquist 1987). The gas and stars in the disk were evolved by means of a hybrid SPH/$N$-body code which is a fully Lagrangian scheme. Dynamic gravitational softening and dynamic smoothing in the gas have been implemented, which resulted in approximately the same accuracy in the smoothed quantities everywhere, or, in other words, to a dynamic spatial resolution. The two-component galactic disk was embedded in a spherically-symmetric halo plus bulge potential.

## 4.1 Initial Conditions and Star Formation

The following important point influenced our way of modeling: within 2-3 kpc from the rotation axis, the disk cannot be assumed geometrically thin and hence the gas and stars cannot be studied by invoking planar motions only. This means that a fully 3D gravitational potential is needed in order to reconstruct the motion in and above the galactic plane. However, the uncertainties in the density distribution in NGC 4321 do not warrant the solution of the Poisson equation. We also decided against running a spectrum of models performing a parameter search for the gas flow in the frozen stellar background. The reason for this is that the amount of gas expected to accumulate within this region is sufficiently large to strongly modify the stellar dynamics there. Freezing of the stellar disk, therefore, would result in unrealistic models as it completely ignores the feedback of the gas on the stars (Shlosman & Noguchi 1993). Instead, we have analyzed the gas dynamics in a 3D model sufficiently similar to NGC 4321 but by no means identical. Only gross features which show robust behavior have been studied. We aim at reproducing (i) the offset shocks in the large scale stellar bar which are delineated by a chain of H II regions and dust lanes; (ii) the nuclear ring-like structure of SF having four distinct maxima; (iii) the skewing of isophotes from the PA of the bar to that of the "ring", both inside and outside the "ring" (as seen in Figs. 2 and 8); and (iv) the approximate alignment between the outer and inner bar structures.

An axisymmetric model of a galactic disk was constructed which is dynamically unstable and develops a stellar bar in the process of evolution. Different parameters of this system have been chosen such as to emphasize the major features of the observed rotation curve in the central few kpc (Arsenault *et al.* 1988; Knapen *et al.* 1993a), *i.e.* the steep velocity gradient within the inner $\sim 400$ pc followed by a relatively slow growth up to $\sim 8-9$ kpc. The initial mass distribution is that of Miyamoto-Nagai (*e.g.* Miyamoto, Satoh & Ohashi 1980) and consists of three components: bulge, halo and disk. A set of hydrodynamical equations coupled to Poisson equation (Miyamoto *et al.* 1980) has been solved to obtain the 3D gravitational potential, the density and the velocity dispersion for each component. The parameters of the three models, Q0 (pure stellar), Q1 (stars+gas, no SF) and Q2 (with SF), are given in Table 1. The bulge and the halo are assumed to be frozen but the disk components, stars and gas, are evolved self-consistently. The halo mass fraction within the circumnuclear region of 2-3 kpc is negligible, however it does govern the global stability of the disk as well as the size and other relevant parameters of the forming bar. It is, therefore, of prime importance in the model. The units of mass, distance and time are $[M] = 1.6 \times 10^{11}$ M$_\odot$, $[R] = 10$ kpc and $[\tau] = 3.75 \times 10^7$ yr. The isothermal equation of state is used for the gas with a temperature of $10^4$ K, except for the regions of SF which are treated differently (see below). The disk contains 32,768 collisionless particles (stars) and 8,192 collisional SPH particles (gas). For the collisionless particles the gravitational softening is kept constant, $\epsilon_* = 200$ pc, while for the SPH particles the softening length is set equal to the smoothing length with a lower resolution limit of $\epsilon_{min,gas} = 200$ pc for models without SF. Models with SF do not require any limit on softening. Finally, the gas is introduced at time $\tau = 0$ by replacing some of the stellar particles by the SPH particles within $\pm 100$ pc from the equatorial plane and inside the radius of 10 kpc, with a radial density distribution identical to that of the stars. We require that the gas amounts to 3% of the total mass within 10 kpc.

The phenomenological approach to treat the effects of SF can be summarized as follows (Heller & Shlosman 1994): (i) a gas particle is considered to undergo "star formation" if the gas is Jeans unstable, if it participates in a converging flow, and if the density in the gas exceeds a critical one, typically $\sim 20$ $M_\odot$ pc$^{-3}$; (ii) only massive OB-type stars are formed leaving no remnants; (iii) the OB stars affect the gas by means of their line-driven winds, $10^{51}$ ergs per $\sim 10^6$ yrs of main sequence evolution, and by supernovae (SN), $10^{51}$ ergs per $\sim 10^4$ yrs. The deposited energy-to-gas 'turbulent' motion conversion efficiency is assumed $\xi \sim 0.05$. The rest of the energy is radiated away instantly (isothermal equation of state). Effects due to the initial mass function (IMF) are ignored which is partly justified by observational evidence that nuclear starbursts have an IMF skewed towards massive stars (*e.g.* Rieke *et al.* 1980; Wright *et al.* 1988; Scalo 1989; Rieke *et al.* 1993).

## 4.2 Modeling Evolution

A stellar bar, extending up to corotation, forms after about one rotation period in the disk in all three models. The steep velocity gradient due to the centrally concentrated mass distribution ensures that the bar is short and of moderate strength. It is capable of driving weak spiral arms only, which in the Q0 model disappear quickly due to the heating of the stellar "fluid". As we are not interested in the transient behavior of the system during the bar instability, we omit this from our discussion (see Heller & Shlosman 1994). The stellar component reaches a steady state after a couple of rotations, which is verified by measuring the strength $q$ of the bar defined here as the maximum ratio of the sum of $m = 2, 4, 6$ and 8 Fourier components to $m = 0$. We measure $q \sim 0.4$ for Q0, and $q \sim 0.3$ for Q1 and Q2 models. Q1 shows the largest fluctuations around this value of $q$. The bar's pattern speed has stabilized at $\Omega_b \sim 2.1$ (for Q0) with a slight decline during the simulation time of $2 \times 10^9$ yrs. Models with gas have a slightly higher bar pattern speed ($\Omega_b \sim 2.6$ for Q1 and Q2), as noted in Heller & Shlosman (1994). The gas component never reaches a true steady state although its evolution slows down substantially after the initial transient phase.

The chosen mass distribution leads to four main *linear* resonances: the outer Lindblad resonance (OLR), the corotation resonance (CR) and the two inner Lindblad resonances — the outer ILR (hereafter OILR) and the inner ILR (hereafter IILR). The latter resonances are the most important ones for our discussion as they are located within the central region, at $r \sim 220$ pc (IILR) and $\sim 1.8$ kpc (OILR) for Q0, and slightly closer to each other in Q1 and Q2, at 240 pc and 1.4 kpc respectively. The CR is at $\sim 4$ kpc. Fig. 9 shows the inner resonances at the time $\tau = 20$. Note that these numbers are based on the assumption that epicyclic approximation is valid, which means a nearly axisymmetric distribution of material in the disk when the orbits of individual stars are more or less circular. As the bar grows and the amplitude of noncircular motion increases, those results become more erroneous. We use fully nonlinear analysis below to validate the positions of the ILRs.

### 4.2.1 Orbit Analysis, Resonances and Gas Dynamics without Star Formation

In order to understand the specific features of gas flow, we have analyzed the dominant families of periodic orbits in the *NGC 4321-like* gravitational potentials of three numerical models. Our analysis is limited to orbits confined to the equatorial plane. Orbits out of the plane are adressed separately (Heller & Shlosman 1995). This approach was pioneered by Contopoulous & Papayannopoulos (1980), Athanassoula *et al.* (1983), Pfenniger (1984) and Sparke & Sellwood (1987), among others (see also Sellwood & Wilkinson 1993 for a review). We first provide some theoretical background to the orbit analysis and subsequently discuss the evolution of numerical models. In short, we have symmetrized the gravitational potential in the plane with respect to the bar's major axis and searched for the $x_2$-type of periodic orbits. These orbits are located between the IILR and OILR and are aligned perpendicularly to the stellar bar. The $x_2$ orbits exist only between the ILRs and in this way serve as their signature and as extension of linear resonances into the nonlinear domain. The reason for the sharp transition between $x_1$ and $x_2$ orbits can be understood within the framework of forced oscillations (*e.g.* Landau & Lifshitz 1969). Using the epicyclic approximation it was shown already by Sanders & Huntley (1976) that the response phase of stellar orbits to the bar potential changes abruptly by 90° at each resonance which leads to the appearance of the $x_2$ orbits between the ILRs and a coexistence of both families in the resonance neighborhood.

While stars can populate different families of orbits within the same region, *e.g.* between the ILRs, this is not true of gas, which in fact does not experience a true resonance at the ILRs due to the damping term in the linearized equation of radial oscillations (Huntley 1977; Huntley, Sanders & Roberts 1978; Binney & Tremaine 1987). Instead, a gradual bending of periodic orbits is predicted which leads to the "crowding of orbits" in the neighborhood of the ILRs and to the formation of trailing spiral arms between the CR and the OILR (crossing the OILR), and of leading spirals across the IILR (Sanders & Huntley 1976; Huntley 1977; Huntley, Sanders & Roberts 1978). The gas orbits hence are aligned with the stellar bar close to the CR and are skewed gradually in the leading direction at smaller radii in the vicinity of the OILR. This phase shift in the gas response should reach a maximum somewhere between the ILRs (close to 90° if the ILRs are separated enough) and then start to decrease. The gas orbits are again aligned with the stellar bar inside the IILR. "Crowding" of streamlines in a highly supersonic ISM flow leads naturally to shocks, and the misalignment of gas and stellar responses — to gravitational torques and to the gas inflow within the CR. We note also that the appearance of shocks invalidates the epicyclic approximation for the gas flow and many of its conclusions, *e.g.* such as signs of gravitational torques — the gas response is highly nonlinear.

Figs. 10a,b represent the anatomy of simple periodic orbits in the full gravitational potential of models Q0 and Q1 at time $\tau = 20$ when both gas and stars reached the approximate steady state. Much of the innermost morphology in the models can be understood using these diagrams, although because of shocks and dissipation in the gas these orbits are only remotely related to the gas motion. The main families $x_1, x_2, x_3$ (prograde) and $x_4$ (retrograde) are marked. Typical orbits which belong to $x_1$ and $x_2$ families are seen in Fig. 11 for different Jacobi 'energies' $E_J$ and the domain of $x_2$ is evident (see Binney & Tremaine 1987 for definition of $E_J$). The $x_2$ orbits for the Q1 model are confined to $E_J \approx$ (-5.24) — (-4.21) at $\tau = 20$

and change little with time. We thus adopt the above limits as nonlinear ILRs in the $E_J$ space. The corresponding semimajor orbital axes (Figs. 10, 11) are $\sim 500$ pc and $\sim 1.3$ kpc. In comparison with the linear ILRs (Fig. 9), this shows that the region between resonances in the nonlinear regime is reduced for all models.

The responses of stellar and gas components in the models are as following: in the pure stellar model (Q0) all isodensity contours within the bar are aligned with the bar. This means that although $x_2$ orbits, as shown in Fig. 11, are allowed and the ILRs are present, most of the stars remain on $x_1$ orbits — a sign that the stellar population in the model is too hot to be trapped. This response is more complex in the presence of gas (model Q1): the gas is losing its angular momentum to the stellar bar between the CR and OILR and falls through toward $x_2$ orbits due to gravitational torques. Some of the gas inside the IILR acquires angular momentum, flows out across this resonance and settles on the innermost $x_2$ orbits. Close to the CR the gas response is parallel to the stellar bar, but further inwards the gas is responding with increasing phase shift producing a pair of offset shocks (Fig. 12). These shocks are curved and concave towards the bar major axis in agreement with 2D numerical simulations of gas flows in analytical bars which have clearly shown that the shock curvature is a function of the bar's strength (*e.g.* Athanassoula 1992). We note the difference between straight shocks which accompany strong bars and curved shocks, produced in our simulations, more typical of weak and moderate bars. We return to this issue in Sections 4.2.2 and 5.

Between the ILRs, the gas response leads the stellar bar by more than 45° — a phase shift which gradually decreases further inwards. The kinematics of this region is clearly dominated by two pairs of shocks: a tightly-wound trailing pair (frequently called 'nuclear spirals' by observers) discussed above and a leading one which crosses the IILR. Figure 13, at times $\tau = 15$ and 17, illustrates both the formation of the shock system and the intimate relation between shocks. In fact, much of the time one can talk about a single pair of shocks which exhibits a pronounced cuspy feature (caustic) between the ILRs, where the leading arms switch to the trailing ones. So, to the extent that these shocks delineate spiral arms, we observe a pseudo-ring made out of a pair of tightly wound spirals between the ILRs. This particular morphology results in the gas accumulating within two elliptical rings between the ILRs due to *shock focusing* (Figs. 12–14). The outer ring forms from the shocked gas which flows along the loci of the offset *trailing* shocks crossing the OILR as they curl around the central region. This ring is positioned deep inside the OILR, roughly speaking the inflowing gas settles on the outer nonintersecting $x_2$ orbits. The inner ring forms from the shocked gas that flows out along the loci of two *leading* shocks. The major axis of this ring is defined by the cuspy feature described above, and extends initially outside the IILR (Figs. 13 and 14ab, *e.g.* before $\tau \sim 22$). The outer ring develops first but both grow at approximately the same median rate $\sim 0.5$ $M_\odot$ yr$^{-1}$.

Both rings persist for the duration of the simulation and appear to be initially closely aligned. While the angle between the stellar bar and the outer ring slightly increases with time from $\sim 70-75°$ to $\sim 80-85°$, the relative PA of the inner ring evolves in the opposite direction. This is illustrated by the orientation of shocks, trailing and leading, in the vicinity of the rings (Fig. 13), the blown-up figures of the rings, and the velocity field there (Figs. 14a,b). The accumulation of gas in the rings increases the gravitational torques they experience from the stellar bar. The rings subsequently evolve as to minimize the total torques on the gas, which is the main reason why the outer ring orients itself along the $x_2$ orbits, more or less, while the inner ring, which shrinks towards the IILR with time, stabilizes along the $x_1$ family, leading it by less than $10°$ (Figs. 13 and 14a,b; note the abrupt change in the orientation of the outer ring between $\tau \sim 19-20$). After $\tau \sim 20$, the morphology of the central region persists for the rest of the numerical simulation, $\sim 2 \times 10^9$ yrs, although the shocks alternate in strength and the continuity between them is not so evident as before. The relative orientations of major dynamical constituents are now similar to those observed in NGC 4321. In particular, there are two density enhancements in the resonance region which are located where the large-scale offset shocks and the flow alongside join the outer ring, *i.e.* in the neighborhood of the bar minor axis and downstream from there. An additional pair of density enhancements is associated with the leading shocks in the inner ring which culminate at the apocenters of the flow in the same region, lying slightly ahead of the bar's major axis. Both are depicted in Figs. 13 and 14a,b and correspond to the loci of SF and peaks of the CO distribution in the region, as discussed below. We emphasize that because the inner ring is strongly elliptical, there is a substantial flow slowdown and gas accumulation at the apocenters of circulation within the ring.

Although the rings corotate with the stellar bar throughout these simulations, they represent only a *temporary slowdown* of the gas inflow. The outer ring "leaks" the gas onto the inner ring which contains orbits deeper in the potential well. As the inflowing gas rotates faster than the inner ring, it shocks continuously against the rear of the ring reinforcing the leading shocks there. Becoming more massive, the inner ring shrinks across the IILR ($\tau \gtrsim 22$) and further inwards. At the late stages of evolution the inner ring becomes increasingly gravitationally unstable to fragmentation because of a high surface density (Toomre 1964; Goldreich & Lynden-Bell 1965). However, this evolution may be already affected by the limiting gravitational softening adopted. About a third of the total amount of gas in the disk has been accumulated in the rings by the end of the numerical simulation, $\sim 6 \times 10^8$ M$_\odot$ in the inner and $\sim 10^9$ M$_\odot$ in the outer ring.

### 4.2.2 Effects of Star Formation in the Resonance Region

Owing to many uncertainties in our treatment of SF, the results of this Section should be viewed as qualitative ones. Technically, SF is introducing "turbulent" motion into gas (see Section 4.1) and induces mixing between gas with different angular momentum. This effect is best seen in the ring evolution between the ILRs (model Q2). Rings widen and share a large fraction of the material. At a later time they can be considered merged for all practical purposes and we shall refer to the gas distribution within the OILR as a disk. Although the "hole" in the gas distribution within what was an inner ring in the Q1 model is now closed, the outer boundary of this disk is well defined and related to the outermost $x_2$ orbit. It has a slightly oval shape and is positioned almost at right angles to the stellar bar (as in the Q1 model).

Throughout this paper we have been emphasizing the close relationship between the shocks and the SF within CR. Model Q2 (Fig. 15) illustrates this dependence by showing the shock system and SF within the central region of 1.3 kpc. Not all the regions of shocked gas are gravitationally unstable, as the comparison between the shock and SF maps show.

The SF is clearly delineating two, usually elongated regions where the inflow along the outer shocks encounters the gas between the ILRs. In addition, a double peak dominates the SF in the neighborhood of the IILR, at the position of the cuspy feature, lying slightly ahead of the stellar bar's major axis, as discussed earlier (Figs. 13, 14a,b), where the gas flow is reaching its apocenter of circulation (see Section 4.2.1). All four regions of intense SF correspond to the maxima of dissipation in the gas of the Q1 and Q2 models. The two SF regions on the bar's major axis also show gas accumulation due to being the apocenters of the gas flow in the inner "ring". Large surface density in the gas and velocity minima lead to Jeans instability here. Furthermore, a certain degree of SF asymmetry is seen in the SF map (Fig. 15) of the central region in accordance with the H$\alpha$ observations described in Paper I. A simple explanation to this effect is that although the SF in the model is driven by a bar-induced density wave, Jeans instability is a local phenomenon affected by the clumpiness in the gas. This introduces a degree of stochasticity to the distribution of Jeans-unstable regions along the wave.

The flow pattern is strongly perturbed in the vicinity of SF due to the locally induced "turbulence", but reforms again and again. Hence the prevailing morphology inferred from the H$\alpha$ imaging is reproduced and shows a robust behavior. Only at the latest stage of the numerical simulations, as we discuss below, this picture changes due to the overwhelming effect of energy deposition by the massive stars which substantially modify the nature of the gas circulation at the center. The SF between the ILRs, *i.e.* between $0.5 - 1.3$ kpc, reaches the peak rate at $\tau \sim 24 - 28$. Its amplitude is variable to within a factor of $\sim 3$ exhibiting a burst behavior with a typical time scale of $\sim 10^7$ yrs, similar to that found by Heller & Shlosman (1994) and which is due to ability of massive stars to destroy the flow pattern when a sufficient number of them coexist in a particular place.

The evolution of Q1 and Q2 models bifurcates after $\tau \sim 28$ ($\sim 2 \times 10^9$ yrs): in Q2, the mass inflow rate across the IILR peaks up strongly, almost tenfold, practically resulting in the catastrophic loss of the angular momentum and collapse of inter-resonance material towards the central 500 pc with the final product being a "fat" and oval disk, as found by Heller & Shlosman (1994). After $\tau \sim 32$ the relative orientation of this disk is almost exactly parallel to the stellar bar, because the gas settles on the $x_1$ orbits deep within the IILR due to gravitational torque from the stellar bar. The SF now is almost exclusively concentrated within the IILR being peaked at the nucleus and the overall morphology has little resemblance to that of NGC 4321. Based on this numerical model, we conclude that a characteristic time scale for "filtering" the gas across the inner resonance zone is less than $\sim 10^9$ yrs.

# 5  Discussion: NIR Morphology and Underlying Dynamics

With the new multiwavelength high-resolution data we are in a position to perform a detailed dynamical study of the gas flow across the nuclear resonance region. Our insight comes from the ability to follow the dust lanes, the exact locations of massive SF, the major concentrations of neutral gas, and the stellar distribution as seen in the NIR, mostly free of the influence of dust extinction (Paper I). Our first and foremost observational result is the confirmation that NGC 4321 is subject to a global density wave driven by a moderately strong and short stellar bar. A number of features observed in NGC 4321 and discussed previously support the resonance character of the circumnuclear region. However, the following two observations are crucial in this respect. First, the strongly offset dust lanes extending from CR inwards and delineating shocks in the ISM indicate the presence of two ILRs. This is related to the existence of a family of perpendicular ($x_2$) orbits (if only one or no ILR is present, these orbits disappear and the dust lanes are centered along the bar's major axis). Second, the $K$ isophotes are skewed backwards from the PA of the "ring" toward the PA of the stellar bar, both outside and inside the "ring" (Fig. 8). The fact that we see the $K$ isophotes returning to the PA of the bar at smaller radii is indicative of the presence of an IILR there. Theoretically, leading spiral arms are expected in the vicinity of the IILR and we believe that our Figs. 2 and 8, showing the twisting of $K$ isophotes backwards to the PA of the stellar bar, provide the first direct observational indication that such arms indeed exist in the central kpc of NGC 4321, and that the IILR is resolved in this case (see also Paper I).

The existence of a double ILR in NGC 4321, however, does not shed light on the problem of a single/double stellar bar there. Strictly speaking, the almost perfect match between the PAs of the large bar and the inner barlike structure on a scale of a few hundred pc could be a chance alignment, although the probability for this seems to be extremely low. It is the existence of the leading armlets extending from the ends of the inner "bar" further out that is difficult to explain within the framework of two nested bars rotating with different pattern speeds in the same direction. We see no evidence that NGC 4321 possesses gaseous or stellar 'bars within bars' which is a rather short-lived phenomenon and, therefore, a sufficiently rare one (Shlosman, Frank & Begelman 1989; Friedli & Martinet 1993; Combes 1994). Summarizing this indirect observational evidence, together with the results of our numerical modeling, we conclude that NGC 4321 hosts only one stellar bar of $60''$ radial size.

The UV, optical and NIR images provide strikingly different views of the circumnuclear region. At optical wavelengths, it is dominated by a pair of well-defined tightly-wound spiral arms which extend inwards and culminate in the intense bursts of SF. Our $U$, $B$ and H$\alpha$ images (Figs. 1, 5) show the prominent sites of SF, somewhat shielded by the dust in its strongest concentrations. The elongated curved arm segments in these images are aligned along the trailing offset shocks seen in dust. It is remarkable that these arms are so tightly wound that they fill up the available space around the inner "bar", and in contrast to arms which form further out in galactic disks, there is not much room for interarm zones, as is clearly seen in the HST image (Fig. 6). The H$\alpha$ arms are parallel to the dust lanes and partially embedded in dust, although at higher HST resolution the bright H$\alpha$ regions within the arms are seen frequently arranged in chains (spurs) perpendicular to the arms. The strong peaks of SF, H$\alpha$1 to H$\alpha$4, show up in all young star indicators (see also Paper I). As revealed by numerical simulations, the peaks H$\alpha$1 and H$\alpha$2 can be associated with the change of character in the spiral arms — from trailing into leading arms. They show not only evidence for young massive stars, but also coincide with the zones K1 and K2, peaking in the dust lanes which cut through H$\alpha$1 and H$\alpha$2. The zones H$\alpha$3 and H$\alpha$4 are strong in OB star indicators ($U$, $B$, H$\alpha$) as well, but are almost absent in the $K$-band, barely affecting the $K$ isophotes in their neighborhood (Fig. 2). This is further confirmed by the progressive blurring of the arm structure seen in the broad-band images with increasing wavelength. In $I$ there is little arm-interarm contrast, and in $K$ it has essentially disappeared (Fig. 1).

Of the four key SF zones in the nuclear ring, H$\alpha$1 (and K1 — its NIR counterpart) is most notable morphologically for its round form in the visible and in $I - K$, where it is seen to be surrounded by a circular dust pattern. This gives a strong impression of an expanding starburst pushing back a dust shell, and it would be interesting to test this with suitable kinematic observations. The H$\alpha$2 zone is less prominent: based on comparison with photometric models (Paper I) we believe it is in an earlier stage of starburst activity than is H$\alpha$1, and still hidden in the dust. Its K2 counterpart is reassuringly similar to K1 — both lie on a straight line through the nucleus, equidistant from it and have almost identical morphologies and luminosities in $K$, all of which supports the above argument. Based on the inferred morphology, we identify both K1 and K2 with the cuspy feature in our numerical simulations. As in the numerical simulations, it is observed slightly to lead the major axis of the stellar bar.

NIR imaging reveals a $60''$ radial length, moderate strength bar, which contributes significantly to the underlying mass distribution, and is barely detectable at visible wavelengths. The stellar bar is 'disected' by a region around 1.3 kpc where $K$ isophotes have low, $\sim 0.13$, ellipticity and show a maximal departure from the PA of the bar (Fig. 8). The curved dust lanes, as seen in the $I - K$ color index image, and also on the HST image, are indicative of shocks in the gas. They appear strikingly symmetric in $I - K$ (Fig. 4) and seem to be a part of a global density wave induced by the stellar bar. Numerical simulations, ours and elsewhere, indicate that weak and moderate bars, which also have a higher probability of possessing ILRs, have curved shocks along the leading edges of the bar accompanied, as is evident from observations, by a string of H II regions (Athanassoula 1992 and refs. therein). On the other hand, SF seems to be inhibited in straight shocks accompanying strong bars due to the large shear in the postshock region (Athanassoula 1992; Heller & Shlosman 1994).

The ring-like structure is clearly formed by a pair of tightly-wound spirals which are delineated by dust and a string of the H II regions slightly offset from dust lanes. The H$\alpha$ intensity rises steeply after the spiral arms cross the bar's minor axis and approach the major axis along the H$\alpha$1 and H$\alpha$2 arcs. These arms continue further inwards for another 180° along the H$\alpha$4 and H$\alpha$3 arcs, respectively, and terminate at the bar's major axis, at K1 and K2 maxima (Figs. 4, 6). The H$\alpha$ intensity then drops again at around 500 pc which is just *outside* the region of steep velocity gradient. The intensity of CO emission has, apart from the nucleus, four pronounced maxima in the dust lanes penetrating the circumnuclear SF zone. The first two maxima can be found where the dust lanes cross the bar's minor axis (and the ring-like SF zone), while two additional maxima are located on the bar's major axis, at the dust lane crossings.

We estimate the total SF rate within the central kpc at $\sim 0.3$ M$_\odot$ yr$^{-1}$, using the H$\alpha$ luminosity (Kennicutt 1983) although not taking into account the dust obscuration. For a comparison, our numerical simulations provide an inflow rate of $\sim 0.5$ M$_\odot$ yr$^{-1}$, with occasional fluctuations by a factor of $\sim 3 - 5$ around this value. This infall toward the central kpc is clearly driven by the stellar bar. The SN rate in the region can be then inferred at just below 0.01 yr$^{-1}$, assuming that stars more massive than 5 M$_\odot$ become SN. Non-thermal radio emission from SN remnants (SNRs) in the circumnuclear "ring" has been detected by $2''$ resolution VLA observations (Weiler *et al.* 1981), who claimed a spectral index $\sim -0.8$ in the range of $1 - 23$ GHz. The radio map of the central region shows similar morphology to that detected in H$\alpha$ (compare their Fig. 1 with our Fig. 5) and allows us to obtain an independent estimate for the SN rate in the region, following Ulvestad (1982), at $\sim 0.02$ yr$^{-1}$. Taken at face value this would mean that the H$\alpha$ extinction in the starburst zone is around a factor 2–3. Correcting the SF rate in the central kpc for dust obscuration and using the estimated amount of molecular gas there, gives a characteristic consumption time of $\sim 6 \times 10^8$ yrs. At the same time, the SNRs currently dominating the radio emission from the "ring" should be younger than $5 \times 10^7$ yrs (*e.g.* Ulvestad 1982). If mass supply to the "ring" is intrinsically intermittent with a characteristic time of a few$\times 10^7$ yrs — the local orbital time scale, as hinted by numerical simulations (here and in Heller & Shlosman 1994), the current population of SNRs may very well be an indicator of the latest burst of SF activity at the center of NGC 4321. A similar time scale has been confirmed for the nuclear starburst galaxy NGC 7552 (Forbes *et al.* 1994) using population synthesis modeling.

The high-resolution $K$ image presented here is central to our understanding of the inner morphology which, as seen in NIR, is very different from that in visible light, *e.g.* in the V-band HST image or in H$\alpha$. This difference is partially due to the transparency of the ISM at 2.2$\mu$ where effects of dust extinction on the morphology are greatly reduced ($A_K/A_V = 0.11$). While localized traces of extinction are still present in $K$ (Fig. 2), as evident after correcting for dust extinction (Paper I), we conclude that the morphology is primarily determined by emission.

What is the origin of NIR light in the central region of NGC 4321? In the absence of a strong AGN, the $K$ luminosity in a starburst galaxy comes mainly from (1) the main population of stars, which for a Salpeter IMF means red giants (dominant contribution in quiescent galaxies); and from (2) red supergiants. In addition, about 20-30% of the $K$ light can be contributed (in extreme cases) by (1) free-free emission from ionized gas; (2) massive blue stars; and (3) very hot $T > 500$ K dust (*e.g.* Telesco 1994).

The distribution of $K$ light from red giants is expected to follow the overall mass distribution because the number of giants is, under very general assumptions, proportional to the total mass in stars. We have noted before that red and NIR isophotes are aligned with the stellar bar close to CR and inside the circumnuclear "ring", while gradually skewed towards the PA of the "ring" at intermediate radii. This *gradual* skewing of the PAs of $K$ isophotes is difficult to interpret in terms of an old stellar population, which should be either aligned with the bar ($x_1$ orbits), or perpendicular to it ($x_2$ orbits). However, twisting of isophotes can be explained invoking a non-negligible contribution from red supergiants, *i.e.* from young stars, which trace the gas flow. If sufficient gas is trapped on these orbits, its gravitational attraction may also influence the main stellar population (Shaw *et al.* 1993), depending on the amplitudes of the stellar dispersion velocities, *i.e.* the temperature of the stellar "fluid". The newborn stars move away from their places of origin because of azimuthal (phase) mixing and because of heating-up in a few orbital times, some $10^8$ yrs in the present case, while more than $10^9$ yrs would be needed for the $K$ light to be dominated by the low-mass giants. A 30 M$_\odot$ O star evolves towards the red supergiant phase in a shorter time ($\sim 5 \times 10^6$ yrs), and should generally follow the original gas orbits, thus preserving the $K$ isophote twisting. At the same time, the more delicate spiral structure will be largely erased in the NIR.

Supergiants have been found to contribute much of the NIR light in starburst galaxies

(*e.g.* Rieke *et al.* 1980; Scoville *et al.* 1988; Forbes *et al.* 1992; Forbes *et al.* 1994; Telesco 1994). This trend seems to be present within the center of NGC 4321 as well: after correcting the dynamical mass (obtained from the measured rotation velocities, Arsenault *et al.* 1988; Knapen *et al.* 1993a) by the observed mass of molecular gas (from Cepa *et al.* 1992), the estimated stellar mass-to-$K$ luminosity ratio $M_*/L_K$ is $\sim 0.3 - 0.4$ M$_\odot$/L$_\odot$ inside 1 kpc, compared to $M_*/L_K > 0.7$ M$_\odot$/L$_\odot$ outside the ring, the latter being typical of a giant-dominated stellar population (Worthey 1994). This indicates that the $K$-emission from the circumnuclear region contains a significant fraction of light from a young stellar population, mixed with light from a normal old population characteristic of stellar bars.

Is this consistent with the appearance of the K1/2 peaks in the NIR? If red supergiants are causing the $K$ peaks, they must be dynamically confined for times comparable to $\sim 10^7$ yrs which correspond to the evolution of an OB star to the supergiant phase. Otherwise a star born from a gas moving with a relative velocity of $\sim 30$ km s$^{-1}$ to the spiral pattern would be found at a distance of $\sim 300 - 400$ pc from its birthplace at the time it becomes a red supergiant. This would erase any spiral features as well as the two prominent peaks in $K$ in the central region unless the latter are gravitationally confined. Hence the K1/2 regions may represent aging starbursts. Numerical simulations described in the previous Section show that the gas is slowed down substantially in the vicinity of the cuspy features which we have associated with the K1/2 peaks. Kinematically, the K1 and K2 regions are especially favorable for gas accumulation, for Jeans instabilities in the gas and for SF (Fig. 15).

Alternatively, young blue massive stars and not red supergiants could dominate the $K$ band continuum of the K1/2 regions with an additional contribution from very hot dust (which is heated directly by stellar radiation) and from the free-free emission in the ionized hydrogen and helium. Within this framework, K1/2 are giant SF complexes triggered and *maintained* by the global gas response to the stirring action of the large stellar bar, which again would explain their remarkable symmetry with respect to the nucleus and the bar itself. Indeed, the existence of some hot massive stars can be inferred directly in K1 through H$\alpha$ emission, and all four SF regions reveal similar colors in a $U - V$ vs. $V - K$ diagram (Paper I).

To estimate the contribution from young massive stars, accompanied by free-free and hot dust emission from their H$\alpha$ regions, NIR colors are necessary. We note that similar estimates have been performed for *e.g.* NGC 3310 by Telesco & Gatley (1984) who found that around $10 - 20\%$ of NIR originates in the blue stars (see also Telesco 1994) and a comparable contribution from hot dust was invoked for the NIR stellar bar in NGC 1068 and other starbursts (Thronson *et al.* 1989; Scoville *et al.* 1988). On the other hand, we discard the possibility that collisionally shock heated dust can contribute substantially to the $K$ emission in NGC 4321. Telesco, Dressel & Wolstencroft (1993) argued that collisional heating of dust in SNR shocks is not a viable mechanism to account even for mid-IR emission from a typical starburst, concluding that the dust is primarily heated by light from young massive stars. NGC 4321 being a mild starburst should fall within this category as well.

All four of the H$\alpha$ zones are accompanied by prominent peaks in CO emission which are seen superposed on the dust lanes (Fig. 7). The CO peaks also seem to indicate the regions where the shocks are strongest, as revealed by the numerical modeling (Fig. 15a). If we correctly understand the kinematics of this region, the bulk of CO emission is located slightly upstream from the corresponding H$\alpha$ peaks. Generally in galaxies CO, as a tracer of molecular gas, is found close to dust lanes, because the H$_2$ whose presence stimulates the CO rotational emission is formed catalytically on grain surfaces, and is shielded by dust from dissociation by the UV photons. The emission strength in CO depends on both the column density and the temperature of the molecular clouds detected. The peaks in CO are obviously associated with the dust lanes and with the H$\alpha 1-4$ SF complexes, but we are unable to say whether their location is determined by peaks in the molecular density, maxima of shocks in the neutral gas, the heating of the neutral gas by OB stars, or a combination of these factors. This is an important question which cannot be solved in the present context.

We estimate that H$_2$ constitutes $\sim 10\%$ of the dynamical mass in the central region, using data from Cepa *et al.* (1992). Within a factor of 2, this is typical of other circumnuclear starbursts: *e.g.* NGC 1097 has 7% inside 2 kpc (Gerin, Nakai & Combes 1988), NGC 3504 has 18% within 1 kpc (Kenney, Carlstrom & Young 1993); NGC 3351 — 10% within 500 pc (Devereux, Kenney & Young 1992). It is not quite clear, however, what the distribution of the molecular gas in the central kpc of NGC 4321 is. The CO map presented here (Fig. 7) is not adequate for this purpose because a significant fraction of the gas at scales larger than $20''$ may have escaped the detection in these observations. Although the resolution of $15''$ used by Cepa *et al.* (1992) may just be good enough, the poor sampling in the inner regions makes their profile unsuitable as well. García-Burillo *et al.* (1994) show a profile along the minor axis of NGC 4321 of CO $J = 2 \to 1$ observations (their Fig. 2), at $12''$ resolution, revealing a central depression. If this depression corresponds to the region inside the SF "ring", the two peaks surrounding it are due to enhanced CO emission from the "ring", as indicated also by Canzian's data (see Fig. 7). When confirmed, preferably with single-beam data at higher resolution, this molecular gas distribution would be in agreement with our modeling, revealing a concentration of neutral gas in the SF "ring". Examples of other galaxies which host nuclear molecular rings are NGC 1097 (Gerin, Nakai & Combes 1988), NGC 4314 (Combes *et al.* 1992) and NGC 4736 (Garman & Young 1986), see also Sofue (1991).

Although we have taken the central kpc out of its context in the external framework of the galaxy, it is necessary to refer briefly to this issue. The galactic disk contains two well defined conventional spiral arms beyond the stellar bar region, which are observed in H I and CO, as well as in stellar surface density and H$\alpha$. These arms are in fact connected to the inner spiral structure described here. The H$\alpha$ image (Fig. 5) shows a clear string of H II regions along the bar (in the NW), linked directly with the exterior spiral arms. An overlay of H II regions on the image of NGC 4321 in the Atlas of Galaxies (Sandage & Bedke 1988) or the more modern real-color image (Peletier 1994), confirms that the H$\alpha$ emission here originates along a strong dust lane. A similar effect can be observed on the other (SE) side of the nucleus, although the H II regions are not as conspicuous. The dust lane through the bar however is clearly seen also on the SE side. These dust lanes aligned with massive SF are indicative of a direct coupling between the outer and inner disks. The continuity of the dust lanes was noted by Sandage (1961), and is confirmed by the new data and, supposing they correspond to large scale galactic shocks in the ISM, by our numerical simulations.

# 6 Conclusions

We have obtained a multi-faceted view of the circumnuclear region in NGC 4321 using a number of observational techniques, and have attempted to model it. We have found interlocking morphology, indicative of the presence of two ILRs, and have studied their associated dynamical influence on the gas flow within the CR.

Much of the disk in NGC 4321 is subject to the density wave driven by a $60''$ radius stellar bar of a moderate strength. The bar can be observed directly only when imaged in the NIR. The $K$ isophotes are generally aligned with the bar except for the intermediate region around $18''$ from the nucleus where they take a weakly oval shape oriented at a large angle to the bar and lead its rotation. These isophotes are gradually skewed back towards the bar both inside and outside the $K$ "ring". This particular behavior of the NIR isophotes prompted us to assume and test the hypothesis that the central region is host to a double ILR and study the evolution of gas distribution in the gravitational potential of a NGC 4321-*like* mass distribution, including the SF and its dynamical effects on the gas. We aimed at reproducing the basic morphology within the CR.

The modeling exercise presented here has provided an insight into the stellar and gas dynamics in the barred region of NGC 4321. We have shown that the present morphology there can be explained within the context of gas response to the non-axisymmetric bar potential and to the underlying mass distribution inferred for this galaxy. Our *non*-linear orbit analysis confirmed that a double ILR is associated with this mass distribution and that the $x_2$ family of orbits oriented along the minor axis of the bar exists between the ILRs. We have used the radial extent of $x_2$ orbits to define the width of the resonance region which narrowed it compared to the one inferred from the galactic rotation curve.

Given that the offset dust lanes within the stellar bar represent large scale shocks in the ISM, we have been able to model their formation and evolution for a characteristic time of $\sim 10$ rotation periods, $\sim 2 \times 10^9$ yrs. Neglecting the initial transient, the central morphology appears to be robust, showing four main compression zones, namely two close to the minor axis of the bar (maybe associated with the so-called "twin peaks"), around 1 kpc from the nucleus, where the inward gas flow curls around the circumnuclear region, and two additional compression zones related to the pair of leading shocks which culminate close to the bar's major axis (slightly ahead of it) at the apocenters of gas circulation within the IILR. This is also the region where large-scale trailing shocks turn into leading shocks creating a pair of cuspy features which serve as loci of SF, as was shown explicitly in the model. Kinematically, they correspond to the minima in the velocity field and are Jeans unstable. We identify the latter compression regions with the K1 and K2 starburst zones in the neighborhood of the IILR at $\sim 500$ pc from the center. We also note that the location of the OILR is not characterized by intense SF or gas compression which are located much further downstream. Rather the OILR can be distinguished on the basis of gas kinematics. Based on our orbit analysis, the OILR should be located outside the SF zone, at around 1.3 kpc from the center. This means that most of the SF in the resonance region happens deep inside and close to its inner boundary at the IILR.

We have no evidence that NGC 4321 possesses two separate stellar bars. Based on our NIR surface photometry, the position angles for the outer and inner "bars" are identical to within the error of determination, and they have similar strength. Taken together with the results of our numerical simulations, this leads us to conclude that only one bar exists in this galaxy. Furthermore, we argued that the overall morphology in the NIR is indicative of the presence of young massive stars within the bar in addition to the normal old population. The gradient of $M/L_K$ across the SF "ring" can be explained by invoking photospheric emission from OB stars, as well as free-free and very hot dust emission. At the same time, the twisting of isophotes in the vicinity of the resonance region requires a population of stars which is dynamically young, *i.e.* that follows gas orbits (see also Paper I). We find that red supergiants aged less than a few$\times 10^7$ yrs are the best candidates, which should be verified by future spectroscopical analysis.

NGC 4321 displays a sharp decrease in SF rate within the central 500 pc. The ovally-shaped inner boundary of SF region corresponding to the incoming pair of tightly-wound spirals is positioned just outside the steep velocity gradient zone, around 500 pc, where we expect the IILR. We find that the gas experiences a positive gravitational torque from the bar and moves out, flattening the gas distribution there.

NGC 4321 was selected for this study because it displays a nuclear starburst, but does so in sufficient moderation as not to destroy the evidence for this process by excessively violent circumnuclear activity. It is fair to point out that the actual flows, at least on the scales examined in this paper, so far have not been observed directly. Previously, however, Knapen *et al.* (1993a) did detect the flow around the bar in H I, with non-circular velocities of tens of km s$^{-1}$. This gives us *prima facie* backing for the claim that similar morphology on the smaller scale will lead to similar dynamics and kinematics. Evidently, the flow pattern within the central 1–2 kpc is far more complex then expected, as it involves the flow across the resonance region. The necessary observations to try to detect the flows are currently feasible on $\sim 1''$ scales in the H$\alpha$ using 2D spectral line imaging techniques. Although it is not yet possible to carry out the same kind of observations in H I, since the column densities are too low to permit the required observations on $2'' - 3''$ scale with the VLA, one may be able to detect the gas flows in the molecular component, if this can be observed at the required resolution with a mm interferometer.

The numerical modeling of the inner region in NGC 4321 is complementary to the high resolution multi-frequency observations. It underlines certain aspects of gas and stellar dynamics whose understanding is crucial for explaining the prevailing morphology and the distribution of SF regions. Overall, it exposes the link between the large-scale structure of the host galaxy and its central activity. Of course the numerical results should be approached with caution, especially the inflow rates and the treatment of SF. Throughout this paper, we have emphasized only qualitative similarities between the observations and modeling. Although we have successfully modeled the main features of circumnuclear region, we have no proof that our solution is unique. The bulge used is probably slightly too massive and, therefore, the bar's pattern speed is higher compared to that found by Elmegreen, Elmegreen & Seiden (1989) and Sempere *et al.* (1995). With all this, we find that the evolution of gas in the vicinity of the ILRs is more complex than described by Schwarz (1984) who used an *ad hoc* restitution parameter to follow the dissipative collisions between finite size 'clouds'. We therefore feel that NGC 4321 has served as a useful modeling paradigm, and that the techniques employed here can be developed for further use with individual objects. Our approach offers a set of kinematically testable results. Given the kind of initial observational input we have had: the morphology both in the visible and in the NIR, the mass distribution from the rotation curve, and the locations of SF complexes, the model predicts the directions, locations, and approximate magnitudes of the gas flows and shock pattern within the circumnuclear region. Thus for NGC 4321, but also for other objects, observers using CO interferometers, 2D optical emission line spectrometers, and the VLA, will be able to compare predictions with observations. Interpretation of such observations will never be trivial, if only due to projection effects, but also due to the ability of starbursts to erase the conditions which led to their occurance in the first place.

*Acknowledgements.* We thank Moshe Elitzur, Bruce Elmegreen, Juhan Frank, Robert Sanders and Tom Troland for helpful discussions, and Blaise Canzian for sending us his CO data of NGC 4321 and for his help in using them here. We acknowledge Reynier Peletier for observing NGC 4321 in $K$ for us and for numerous discussions. The WHT is operated on the island of La Palma by the RGO in the Spanish Observatorio del Roque de los Muchachos of the IAC. The UKIRT is operated by the Royal Observatory in Edinburgh on behalf of the SERC. Based on observations made with the NASA/ESA HST, obtained from the data archives at the STScI which is operated by the AURA under NASA contract NAS 5-26555. JHK and JEB are partially supported by the Spanish DGICYT grant PB91-0510. IS acknowledges support from the Gauss Foundation and the IAC (through project P3/86), and thanks Klaus Fricke and John Beckman for hospitality. IS also acknowledges NASA grant NAGW-3839 and continuing support from the University of Kentucky Center for Computational Studies.

# References

Adams, T.F. 1977, ApJS 33, 19
Arsenault, R., Boulesteix, J., Georgelin, Y. & Roy, J.-R. 1988, A&A 200, 29
Athanassoula, E. 1992, MNRAS 259, 345
Athanassoula, E. 1994, in *Mass-Transfer Induced Activity in Galaxies*, Ed. I.Shlosman (Cambridge Univ. Press), p. 143
Athanassoula, E., Bienaymé, O., Martinet, L. & Pfenniger, D. 1983, A&A 127, 349
Balzano, V.A. 1983, ApJ 268, 602
Barnes, J. & Hut, P. 1986, Nature 324, 446
Beckman, J.E., Peletier, R.F., Knapen, J.H., Gentet, L. & Maté, M.J. 1995, in *Opacity of Spiral Discs*, Ed. J. Davies, Kluwer Dordrecht, in press
Begelman, M.C., Blandford, R.D. & Rees, M.J. 1984, RMP 56, 255
Binney, J. & Tremaine, S. 1987, *Galactic Dynamics* (Princeton: Princeton Univ. Press)
Blandford, R.D. & Payne, D.G. 1982, MNRAS 199, 883
Block, D.L., Bertin, G., Grosbøl, P., Moorwood, A.F.M. & Peletier, R.F. 1994, A&A 288, 365
Bosma, A. 1981, AJ 80, 1825
Bressan, A., Chiosi, C. & Fagotto, F. 1994, ApJS 94, 63
Buta, R. 1986a, ApJS 61, 609

Buta, R. 1986b, ApJS 61, 631
Buta, R. & Crocker, D.A. 1993, AJ 105, 1344
Canzian, B. 1990, Ph.D. thesis, California Institute of Technology
Canzian, B. 1993, ApJ 414, 487
Canzian, B. & Allen, R.J. 1995, in preparation
Cepa, J. & Beckman, J.E. 1990, A&ASS 83, 211
Cepa, J., Beckman, J.E., Knapen, J.H., Nakai, N. & Kuno, N. 1992, AJ 103, 429
Combes, F. 1994, in *Mass-Transfer Induced Activity in Galaxies*, Ed. I.Shlosman (Cambridge Univ. Press), p. 170
Combes, F. & Gerin, A. 1985, A&A 150, 327
Combes, F., Gerin, M., Nakai, N., Kawabe, R., & Shaw, M.A. 1992, A&A 259, L27
Contopoulos, G. & Papayannopoulos, Th. 1980, A&A 92, 33
de Jong, R.S. & van der Kruit, P.C. 1994, A&AS, 106, 451
de Vaucouleurs, G. 1963, ApJS 8, 31
de Vaucouleurs, G. 1984, in *The Virgo Cluster*, ed. O.G. Richter & B. Binggeli (Garching: ESO), 413
de Vaucouleurs, G., de Vaucouleurs, A. & Corwin, H.G.Jr. 1976, *Second Reference Catalogue of Bright Galaxies* (RC2), Univ. of Texas Press
Devereux, N.A. 1987, ApJ 323, 91
Devereux, N.A., Kenney, J.D.P. & Young, J.S. AJ 103, 784
Dressel, L.L & Gallagher, J.S. 1994, in *Mass-Transfer Induced Activity in Galaxies*, Ed. I.Shlosman (Cambridge Univ. Press), p. 165
Elmegreen, B.G. 1994, ApJL 425, L73
Elmegreen, D.M. & Elmegreen, B.G. 1984, ApJS 54, 127
Elmegreen, B.G., Elmegreen, D.M. & Seiden, P.E. 1989, ApJ 343, 602
Emmering, R.T., Blandford, R.D. & Shlosman, I. 1992, ApJ 385, 460
Forbes, D.A., Ward, M.J., DePoy, D.L., Boisson, C. & Smith, M. 1992, MNRAS 254, 509
Forbes, D.A., Norris, R.P., Williger, G.M. & Smith, R.C. 1994, AJ 107, 984
Freedman, W. *et al.* 1994, Nature 371, 757
Friedli, D. & Benz, W. 1993, A&A 268, 65
Friedli, D. & Martinet, L. 1993, A&A 277, 27
García-Burillo, S., Sempere, M.J. & Combes, F. 1994, A&A 287, 419
Garman, L.E. & Young, J.S. 1986, A&A 154, 8
Gerin, M., Nakai, N. & Combes, F. 1988, A&A 203, 44
Goldreich, P. & Lynden-Bell, D. 1965, MNRAS 130, 97
Hackwell, J.A. & Schweizer, F. 1983, ApJ 265, 643
Heckman, T.M. 1990, in IAU Coll. 124 on *Paired & Interacting Galaxies*, Eds. J.W.Sulentic *et al.* (NASA CP-3098), p. 359
Heller, C.H. & Shlosman, I. 1994, ApJ 424, 84
Heller, C.H. & Shlosman, I. 1995, ApJ, submitted
Hernquist, L. 1987, ApJS 64, 715
Hummel, E. 1981, A&A 93, 93
Huntley, J.M. 1977, Ph.D. thesis, University of Virginia
Huntley, J.M. 1980, ApJ 238, 524
Huntley, J.M. Sanders, R.H. & Roberts, W.W.Jr. 1978, ApJ 221, 521
Ishizuki, S., Kawabe, R., Ishiguro, M., Okumura, K.S., Kasuga, T., Chikada, Y. & Takashi,

K. 1990, Nature 334, 224
Jansen, R.A., Knapen, J.H., Beckman, J.E., Peletier, R.F. & Hes, R. 1994, MNRAS 270, 373
Kenney, J.D.P. & Young, J.S. 1988, ApJS 66, 261
Kenney, J.D.P., Carlstrom, J.E. & Young, J.S. 1993, ApJ 418, 687
Kenney, J.D.P., Wilson, C.D., Scoville, N.Z., Devereux, N.A. & Young, J.S. 1992, ApJL 395, L79
Kennicutt, R.C. 1983, ApJ 272, 54
Kennicutt, R.C. 1994, in *Mass-Transfer Induced Activity in Galaxies*, ed. I. Shlosman (Cambridge: Cambridge Univ. Press), p. 131
Kennicutt, R.C., Keel, W.C. and Blaha, C.A. 1989, AJ 97, 1022
Knapen, J.H., Hes, R., Beckman, J.E. & Peletier, R.F. 1991, A&A 241, 42
Knapen, J.H., Beckman, J.E., Cepa, J., van der Hulst, J.M. & Rand, R.J. 1992, ApJL 385, L37
Knapen, J.H., Cepa, J., Beckman, J.E., del Rio, M.S. & Pedlar, A. 1993a, ApJ 416, 563
Knapen, J.H., Arnth-Jensen, N., Cepa, J. & Beckman, J.E. 1993b, AJ 106, 56
Knapen, J.H., Beckman, J.E., Shlosman, I., Peletier, R.F., Heller, C.H. & de Jong, R.S. 1995a, ApJL 443, L73 (Paper I)
Knapen, J.H., Beckman, J.E., Cepa, J. & Nakai, N. 1995b, A&A submitted
Kormendy, J. 1982, in *Morphology & Dynamics of Galaxies*, Eds. L.Martinet & M.Mayor (Geneva Observatory), p.115
Kormendy, J. 1994, in IAU Symp. 153 on *Galactic Bulges*, Eds. H.Habing & H.Dejonghe (Dordrecht: Kluwer Acad. Publ.), in press
Landau, L.D. & Lifshitz, E.M. 1969, *Mechanics* (New York: Pergamon Press).
Landolt, A.U. 1983, AJ 88, 439
Lo, K.Y., Cheung, C.R., Phillips, T.G., Scott, S.L. & Woody, D.P. 1987, ApJ 312, 574
Martin, P. & Roy, J.-R. 1994, ApJ 424, 599
Meixner, M., Blitz, L., Puchalsky, R., Wright, M. & Heckman, T. 1990, ApJ 354, 158
Miyamoto, M., Satoh, C. & Ohashi, M. 1980, Ap. Space Sci. 67, 147
Moles, M., Márquez, I. & Pérez, E. 1995, ApJ 438, 604
Morgan, W.W. 1958, PASP 70, 364
Nakai, N., Hayashi, M., Handa, T., Sosue, Y., Hasegawa, T. & Sasaki, M. 1987, PASJ 39, 685
Norman, C.A. & Silk, J. 1983, ApJ 266, 502
Peletier, R.F. 1994, Spectrum Newsletter 3, 28
Peletier, R.F. & Willner, S.P. 1991, ApJ 382, 382
Pfenniger, D. 1984, A&A 134, 373
Pfenniger, D. & Norman, C.A. 1990, ApJ 363, 391
Pierce, M.J. 1986, AJ 92, 285
Pogge, R.W. 1989, ApJS 71, 433
Prendergast, K.H. 1962, in *Distribution & Motion of ISM in Galaxies*, ed. L. Woltjer (New York: Benjamin), p. 217
Puxley, P.J., Hawarden, T.G. & Mountain, C.M. 1988, MNRAS 231, 465
Rand, R.J. 1995, AJ 109, 2444
Rieke, G.H. & Lebofsky, M.J. 1985, ApJ 288, 618
Rieke, G.H., Lebofsky, M.J., Thompson, R.I., Low, F.J. & Tokunaga, A.T. 1980, ApJ 238, 24

Rieke, G.H., Loken, K, Rieke, M.J. & Tamblyn, P. 1993, ApJ 412, 99
Rix, H.-W. & Rieke, M.J. 1993, ApJ 418, 123
Sandage, A. 1961, *The Hubble Atlas of Galaxies* (Washington: Carnegie Institution of Washington)
Sandage, A. & Bedke, J. 1988, *Atlas of Galaxies*, NASA, Washington DC
Sanders, R.H., & Huntley, J.M. 1976, ApJ 209, 53
Scalo, J.M. 1989, in *Windows on Galaxies*, Eds. A.Renzini *et al.*, (Dordrecht: Kluwer Acad. Publ.)
Schwarz, M.P. 1984, MNRAS 209, 93
Scoville, N.Z., Hibbard, J.E., Yun, M.S. & van Gorkom, J.H. 1994, in *Mass-Transfer Induced Activity in Galaxies*, ed. I. Shlosman (Cambridge: Cambridge Univ. Press), p. 191
Scoville, N.Z., Matthews, K., Carico, D.P. & Sanders, D.B. 1988, ApJL 327, L61
Sellwood, J.A. & Wilkinson, A. 1993, Rep. Prog. Phys. 56, 173
Sempere, M.J., García-Burillo, S., Combes, F. and Knapen, J.H. 1995, A&A, in press
Sérsic, J.L. & Pastoriza, M. 1967, PASP 79, 152
Shaw, M.A., Combes, F., Axon, D.J. & Wright, G.S. 1993, A&A 273, 31
Shlosman, I. & Noguchi, M. 1993, ApJ 414, 474
Shlosman, I., Begelman, M.C. & Frank, J. 1990, Nature 345, 679
Shlosman, I., Frank, J. & Begelman, M.C. 1988, in Proc. IAU Coll. No. 134 on *Active Galactic Nuclei*, eds. D. Osterbrock & J.Miller (Dordrecht: Kluwer Acad. Publ.), p. 462
Shlosman, I., Frank, J. & Begelman, M.C. 1989, Nature 338, 45
Simkin, S.M., Su, H.J. & Schwarz, M.P. 1980, ApJ 237, 404
Sofue, Y. 1991, PASP 43, 671
Sparke, L.S. & Sellwood, J.A. 1987, MNRAS 227, 653
Taniguchi, Y., Kameya, O., Nakai, N. & Kawara, K. 1990, ApJ 358, 132
Telesco, C.M. 1988, ARAA 26, 343
Telesco, C.M. 1994, in *Infrared Astronomy*, Eds. A.Mampaso *et al.* (Cambridge: Cambridge Univ. Press)
Telesco, C.M. & Decher, R. 1988, ApJ 334, 573
Telesco, C.M. & Gatley, I. 1984, ApJ 284, 557
Telesco, C.M., Dressel, L.L. & Wolstencroft, R.D. 1993, ApJ 414, 120
Telesco, C.M., Campins, H., Joy, M., Dietz, K. & Decher, R. 1991, ApJ, 369, 135
Thronson, H.A. *et al.* 1989, ApJ 343, 158
Toomre, A. 1964, ApJ 139, 1217
Ulvestad, J.S. 1982, ApJ 259, 96
Wang, Z., Scoville, N.Z. & Sanders, D.B. 1991, ApJ 368, 112
Weiler, K.W., van der Hulst, J.M., Sramek, R.A. & Panagia, N. 1981, ApJL 243, L161
Worthey, G. 1994, ApJS 95, 107
Wright, G.S., Joseph, R.D., Robertson, N.A., James, P.A. & Meikle, W.P.S. 1988, MNRAS 233, 1
Young, J.S., Xie, S., Kenney, J.D.P. & Rice, W.L., 1989, ApJS 70, 699
Zaritsky, D., Rix, H.-W. & Rieke, M.J. 1993, Nat 364, 313

# 7 Figure Captions

**Figure 1.** (Plate 1) Panel of images of the central region in NGC 4321 in six different optical and NIR passbands, $U$ (a., upper left), $B$ (b., upper right), $V$ (c., middle left), $R$ (d., middle right), $I$ (e., lower left) and $K$ (f., lower right). Contours are shown overlaid on a grey-scale representation of the same image in each panel, where darker shades represent lower intensities. Contour intervals are 0.5 mag in all panels, ranging from 20.5 to 17.5 ($U$), 21.0 to 17.5 ($B$), 20.0 to 16.5 ($V$), 19.5 to 16.0 ($R$), 19.0 to 15.0 ($I$) and 17.0 to 13.5 ($K$). The spatial resolution is $0.''8 \pm 0.''05$ in all panels. Area shown and scale are the same in all panels, and also in all subsequent Figures, unless stated otherwise. A bright cosmic ray hit stands out clearly toward the SW in the $B$-band image.

**Figure 2.** Contour representation of the NIR $K$-band image overlaid on a grey scale representation of the same image. Dashed line indicates the position angle of the *large-scale* bar (not shown in this Figure), at PA=107°. K1 and K2 indicate "hot spots" in $K$, further discussed in the text. Contour levels are 16.7, 16.3, 15.9, 15.6, 15.4, 15.2, 15.0, 14.8, 14.6, 14.2, 13.6 and 13.0 $K$-mag.

**Figure 3.** (Plate 2) Grey scale representation of $U - I$ color index image. Darker shades mean redder colors (relatively more $I$ emission). Grey scales vary from $-0.5$ to 3 magnitudes.

**Figure 4.** $I - K$ color index image, where darker shades indicate redder colors or relatively more $K$ emission, which can be interpreted as higher dust extinction. Grey scales are from 1.4 (light) to 3.1 (dark) magnitudes.

**Figure 5.** Continuum-subtracted H$\alpha$ image of the central $3'$ of NGC 4321 showing part of the disk which includes the $60''$ large-scale bar. Effective spatial resolution is $< 1''$. The inset shows an amplification of the circumnuclear region (labels are in $''$; resolution $0.''8$ ), with emitting regions H$\alpha$1, H$\alpha$2, H$\alpha$3 and H$\alpha$4 indicated. The dotted lines are representative $H$-band contours outlining the outer. large-scale bar ($H$ image from Peletier & Willner 1991, kindly supplied by R.F. Peletier). Contour levels for the H$\alpha$ image are 0.7, 2.1, 6.3, 12.6 and $25.2 \times 10^{36}$ erg s$^{-1}$, and (inset) 0.7, 4.1, 2.8, 5.6, 8.4, 11.2, 16.8 and $22.4 \times 10^{36}$ erg s$^{-1}$.

**Figure 6.** Overlay of our $K$ band image (contours as in Fig. 2) on an image in the visible taken with the HST (grey scales, lighter shades in the Figure indicating stronger emission). The HST image has been degraded but its resolution is still superior to that of the $K$-band image.

**Figure 7.** Overlay of a contour representation of an interferometric map of $^{12}$CO $J = 1 \to 0$ emission (courtesy B. Canzian) on our $I - K$ color index map (grey scales). CO contours are at 22, 35, 47, 58, 70, 81 and 93% of peak intensity. $I - K$ greys range from 1.4 (light) to 3.1 magnitudes (dark). The beam size for the CO observations is indicated. Note that all peaks seen in CO (except the central one) coincide with dust lanes, which show up dark in the $I - K$ map.

**Figure 8.** Run against radius of three parameters describing the red surface brightness distribution in NGC 4321, as measured from $I$ (filled dots) and $K$ (crosses) images using ellipse-fitting. Values are plotted up to $R = 35''$ as derived from $K$ and for $10'' < R < 75''$

as derived from $I$. *Upper panel:* surface brightness, in $K$-mag arcsec$^{-2}$ (scale on left side) and $I$-mag arcsec$^{-2}$ (scale on right side). *Middle panel:* position angle, in degrees, measured N over E. *Lower panel:* ellipticity (defined as $1 - b/a$, with $b/a$ the axis ratio). Scale on the right side shows the corresponding angle of inclination, in case the measured ellipticity were due solely to seeing an intrinsically round object projected on the plane of the sky.

**Figure 9.** Linear resonances for the Q0, Q1 and Q2 models at $\tau = 20$. The horizontal lines correspond to the bar pattern speeds in these models: dashed line for the Q0 model and dotted line for the Q1 and Q2 models.

**Figure 10.** Characteristic diagrams of orbit families in the rotating frame of the bar at $\tau = 20$ in *(a)*. The Q0 model and *(b)*. The Q1 model. $y_0$ gives the bar minor axis crossing by individual orbits. The dashed line shows the zero velocity curve (ZVC). Only the main families inside 3 kpc are shown.

**Figure 11.** A representative set of simple periodic orbits in the gravitational potential of the Q0 model. The stellar bar is elongated along the $x$-axis. The $x_1$ family of orbits extends between the OILR and CR, and between the center and the IILR. The $x_2$ family is limited to between the ILRs. Note that the linear analysis (Fig. 9) shows a wider range allowed for $x_2$.

**Figure 12.** *Left:* Logarithmic grey scale map of shock dissipation in the Q1 model emphasizing the large-scale offset shocks between the CR and the OILR at $\tau = 17.4$ (ring formation) and $\tau = 20.0$ (steady state). Each frame is 8 kpc across. The grey level is given by the time derivative of the nonadiabatic component of internal energy. The rings are "overexposed". The gas flow is anti-clockwise and the stellar bar is horizontal. *Right:* Gas flow in the bar's frame of reference which corresponds to the shock pattern on the left. Vector size is scaled with the velocity. Only half of the particles are shown. Note the blobby character of dissipation in the gas. The sharp break in the spiral arms occurs approximately at CR ($\tau = 20$).

**Figure 13.** Logarithmic grey scale map of shock dissipation in the Q1 model emphasizing the evolution of shocks inside the OILR. Time is given in the upper left corners. Each frame is 2.6 kpc across. The grey level is given by the viscous dissipation rate. The gas flow is anti-clockwise and the stellar bar is horizontal.

**Figure 14.** *(a).* Logarithmic gas isodensity contours in the circumnuclear region (Q1 model) shown in Fig. 13. *(b).* Gas flow in the region shown in Figs. 13 and 14a, in the bar's frame of reference. Vector size is scaled with the velocity. Only third of the particles are shown.

**Figure 15.** *Left:* Logarithmic grey scale map of shock dissipation in the circumnuclear region (Q2 model). Time is given in the upper left corners. The gas flow is anti-clockwise and the stellar bar is horizontal. Each frame is 2.6 kpc across. *Right:* SF map corresponding to the region shown on the left.

Table 1: Miyamoto-Nagai model parameters for the halo, bulge, and disk. The components are defined by their mass $a$, vertical scale-height $c$, and horizontal scale $b+c$.

| Model Parameters | | | |
|---|---|---|---|
| | a | b | c |
| Halo | 3.041 | 0.000 | 1.800 |
| Bulge | 0.105 | 0.000 | 0.045 |
| Disk | 0.991 | 0.500 | 0.020 |